\documentclass[a4paper,12pt]{article}

\usepackage{amssymb,amsmath,mathbbol,mathrsfs}
\usepackage[usenames,dvipsnames]{color}
\usepackage{hyperref}
\usepackage{xcolor}
\usepackage{stmaryrd}
\usepackage{authblk}
\usepackage{framed}
\usepackage{empheq}
\usepackage{slashed}
\usepackage{hyphenat}
\usepackage{cite}
\usepackage{apacite}
\usepackage{natbib}

\usepackage{marginnote}    

\usepackage[left=.8in,right=.8in,top=.8in,bottom=.8in]{geometry}                  

\usepackage{sectsty} 
\allsectionsfont{\sffamily\mdseries\upshape} 
\usepackage{tocloft}

\makeatletter
\renewenvironment{abstract}{%
    \if@twocolumn
      \section*{\abstractname}%
    \else 
      \begin{center}%
        {\bfseries\sffamily\abstractname\vspace{\z@}}
      \end{center}%
      \quotation
    \fi}
    {\if@twocolumn\else\endquotation\fi}
\makeatother

\numberwithin{equation}{section}
5

\hypersetup{
	colorlinks=true,         
	linkcolor=MidnightBlue,          
	citecolor=BrickRed,        
	urlcolor=MidnightBlue            
}

\newcommand{\tocite}[1]{[{\color{magenta}\texttt{#1}}]}
\newcommand{\be}{\begin{equation}}
\newcommand{\ee}{\end{equation}}

\renewcommand{\d}{{\mathrm{d}}}
\newcommand{\D}{{\mathrm{D}}}

\newcommand{\SU}{{\mathrm{SU}}}

\newcommand{\G}{{\mathcal{G}}}

\newcommand{\pp}{{\partial}}

\renewcommand{\bar}{\overline}

\newcommand{\WZ}{\mathsf{WZ}}
\newcommand{\wz}{\mathsf{wz}}
\newcommand{\cs}{\mathsf{cs}}
\newcommand{\CS}{\mathsf{CS}}

\newcommand{\ch}{\mathsf{ch}}
\newcommand{\Ch}{\mathsf{Ch}}
\renewcommand{\t}{\mathfrak{t}}

\newcommand{\cint}{{\int\kern-.87em{<}}}
\newcommand{\sint}{{\int\kern-.75em{\sim}}}
\newcommand{\fint}{{\int\kern-1.00em{\int}}}

\newcommand{\bb}{\mathbb}

\newcommand{\tr}{\text{tr}}

\let\oldmarginpar\marginpar
\renewcommand\marginpar[1]{\oldmarginpar{\color{red}\raggedright\footnotesize #1}}

\title{Large gauge transformations, gauge invariance, and the QCD $\theta_{\text{\tiny YM}}$-term}
\author{Henrique Gomes\footnote{University of Cambridge, Trinity College, CB2 1TQ, United Kingdom; \href{mailto:gomes.ha@gmail.com}{gomes.ha@gmail.com}}~ and Aldo Riello\footnote{Perimeter Institute for Theoretical Physics, EC1R 4UP Canada; \href{mailto:ariello@pitp.com}{ariello@pitp.com}}  \\\it }
\begin{document}
\maketitle
\abstract{ The eliminative view of gauge degrees of freedom---the view that they arise solely from descriptive redundancy and are therefore eliminable from the theory---is a lively topic of debate in the philosophy of physics. 
 Recent work attempts to leverage properties of the QCD $\theta_{\text{\tiny YM}}$-term to provide a novel  argument against the eliminative view. The  argument is based on the claim that the QCD $\theta_{\text{\tiny YM}}$-term changes under ``large'' gauge transformations. 
 Here we  review geometrical propositions about  fiber bundles  that unequivocally falsify these claims: the $\theta_{\text{\tiny YM}}$-term  encodes topological features of the fiber bundle used to represent gauge degrees of freedom, but it is \textit{fully} gauge-invariant.  
Nonetheless,  within the essentially classical viewpoint pursued here, the physical role of the $\theta_{\text{\tiny YM}}$-term shows the physical importance of bundle topology (or superpositions thereof) and thus weighs against (a naive) eliminativism.
}

\section{Introduction}
Modern philosophers take seriously the ontological status of fields. But what they usually have in mind are relatively concrete entities, such as the electric and magnetic fields, and not the  elusive gauge fields, such as the electromagnetic potential.  How then, to classify ``gauge'' degrees of freedom? Do these have an ontological significance similar to  electric and magnetic fields, or are they only a notational convenience, born of a redundancy in our representations of the world? In the words of Earman, are gauge degrees of freedom only ``redundant descriptive fluff'' \citep{Earman_sym}?

The eliminativist view of gauge degrees of freedom advocates not only that gauge degrees of freedom are redundant, but that they are also eliminable. One proponent of eliminativism within the philosophy of physics community is Richard Healey, whose position is laid out in \citep{Healey_book}. Healey proposes that we should use a different, gauge-invariant basis to describe our physical quantities: non-local, yes, but controllably so; this is called the holonomy-basis.\footnote{Whether one can really write down a theory---an action functional or a Hamiltonian---in terms of holonomies (or Wilson loops) is challenging, to say the least. But we will not pursue this in this paper.
}

The status of gauge degrees of freedom is too large a topic to be reviewed here.  We plan only to analyze a recent argument against the eliminativist view, and show that it is founded on an incorrect mathematical treatment---and it is therefore not tenable in its current form.  In the rest of this section, we introduce the argument and give a prospectus for the paper.

\subsection{The $\theta_{\text{\tiny YM}}$-term}\label{sec:1.1}
In a recent paper, \citet{Dougherty_CP} engages with the details of \citet{Healey_book}'s  eliminativist program  in the context of QCD. Dougherty's first aim is to convince the reader that a $\theta_{\text{\tiny YM}}$-term in the QCD Lagrangian  is  mandatory. 


 In brief, the argument is as follows: the $\theta_{\text{\tiny YM}}$-term is necessary to account for certain experimental facts.  To be more specific: the smallness of the masses of the up and down quarks gives rise to a chiral symmetry, whose effects (a parity doubling of the hadron spectrum, cf. \citep[Sec. 19.10]{WeinbergQFT2}) are not observed in experiments.  This means that this chiral symmetry must be broken somehow. But the spontaneous breaking of this symmetry would generate Goldstone bosons, which are also not observed. Therefore, {\it one must be able to break chiral symmetry without creating Goldstone bosons}. 
 
 A solution is to have the breaking  be effected through an anomaly.\footnote{ This solution, however, might not be appropriate in a non-perturbative treatment. See section \ref{sec:cautionary}.}
  Namely, under chiral transformations (also called a global U$(1)_A$ symmetry), it turns out that the path-integral measure for quark fields  fails to be invariant and rather acquires a phase. Specifically, for a fermion field of flavor $f$,  the chiral symmetry acts by a shift $\psi_f\mapsto \exp(i\gamma_5\alpha_f)\psi_f$ (with $\gamma_5$ the fifth gamma-matrix),  whereas the fermion path-integral transforms as%
 \footnote{This is the standard argument first put forward by Fujikawa (cf. \citet[Sec. 5.2]{Bertlmann}). Now, the $\theta_{\text{\tiny YM}}\text{-term}$ is a functional of the curvature,  $F_{\mu\nu}$, so why does it appear in a change in the measure of purely fermionic degrees of freedom?  In Fujikawa's implementation of a gauge covariant measure, one writes the fermion field in terms of a basis of eigenfunctions of the Dirac operator, $\slashed{\D}$, which includes the gauge-covariant derivative $ \D_\mu = \pp_\mu + A_\mu$, inside it (i.e. $\slashed{\D}=\gamma^\mu \D_\mu$, where $\gamma^\mu$ are the Dirac matrices).  It then turns out that the determinant of the Jacobian under a chiral transformation in this orthonormal basis diverges and needs to be regularized. Fujikawa used a gauge-covariant Gaussian cut-off by insertion of  the operator $\lim_{M\rightarrow\infty}\exp{(-\frac{\slashed{\D}^2}{M^2})}$. Ultimately, the curvature appears through the decomposition: $\slashed{\D}^2=\D_\mu \D^\mu+\frac14[\gamma^\mu, \gamma^\nu]F_{\mu\nu}$. One can choose a gauge-invariant measure, in which case the anomaly is shifted to counterterms (which necessarily fail to satisfy the same invariances of the Lagrangian, \citet[ vol 2, ch.28]{DeWitt_Book2}).}
\be
\mathcal{D}\psi \mathcal{D}\bar\psi\mapsto \exp\left(i2(\theta_{\text{\tiny YM}}\text{-term})\sum_f\alpha_f \right)\mathcal{D}\psi \mathcal{D}\bar\psi,
\ee
where
\be
\theta_{\text{\tiny YM}}\text{-term} =  \frac{1}{8\pi^2} \int \tr( F\wedge F).
\ee
Therefore, according to this argument, mathematical consistency and experimental evidence---the lack of both the relevant Goldstone bosons and of the parity doubling of the hadron spectrum---together would provide support for the physical significance of the $ \theta_{\text{\tiny YM}}$-term. It is here important to stress the role fermions play in making the $\theta_{\text{\tiny YM}}$-term inescapable.

But this is not the end of the story: such a term would be CP-violating and thus gives rise to other questions of observability. The relation between CP-violation and the $\theta_{\text{\tiny YM}}\text{-term}$  is not  directly relevant to the central points of this paper, which is why we will avoid discussing it.\footnote{Briefly, the field redefinitions above---modifying the definitions of the quarks by a chiral transformation---not only shift the coupling constant $\theta$ in front of the $\theta_{\text{\tiny YM}}$-term in the Yang-Mills Lagrangian by $\theta \mapsto \theta + \sum_f \alpha_f$, but also change the mass terms in the Lagrangian density by $m_f\mapsto \exp(i2\alpha_f)m_f$. Since physical quantities cannot be affected by a mere field-redefinition, this means that the only invariant quantity physical systems can depend on is the {\it product} $e^{- i \theta} \prod_f m_f$ (cf. \citep[Sec. 23.6]{WeinbergQFT2}). This product  defines an invariant version of the $\theta$-coupling, called $\bar \theta$. 
Thus, if one flavor of quarks had zero mass, the puzzle would be resolved. That doesn't seem to be the case. Nonetheless, $\bar\theta$ is observationally constrained to be close to zero: the current bound on $\bar\theta$ is $|\bar\theta|<2 \times 10^{-10}$ \tocite{\href{http://pdg.lbl.gov/2019/reviews/rpp2019-rev-conservation-laws.pdf}{particle data group}, see: \href{http://pdg.lbl.gov/2019/reviews/contents_sports.html}{this} for general citation}. The question of theoretical necessity of  the $\bar\theta$-term hinges on important  issues of naturalness and fine-tuning, and, since there is currently experimental reason to believe that it  vanishes, one might feel compelled to explain its observational smallness. That is, what physicists refer to as the ``Strong CP problem''---that Nature conspires to give the CP-violating $\bar\theta$-term a value close to zero---is a real problem that still  lacks an agreed explanation. But Dougherty does not tie his boat to the issue of explanation for the smallness of $\bar\theta$. \label{fnt:theta}}

Having set the broader context for the discussion, we now  very briefly embed within it  Dougherty's criticism of Healey. Before we begin, it should be stated from the outset that our intention in this paper is only to set straight a specific misunderstanding of this criticism: the gauge-invariance properties of the $ \theta_{\text{\tiny YM}}$-term; we will mostly constrain the remit of our discussion accordingly. 

\subsection{Dougherty's criticism}

According to \citet{Dougherty_CP} (cf p.1, 7, 8, 16) the underlying reason for Healey's elimination of the $\theta_{\text{\tiny YM}}$-term is that such a term is only gauge-invariant under gauge transformations that have a particular behaviour at infinity (or at the relevant boundaries). The idea then is that the \textit{non}-eliminativist would be comfortable in separating the  wheat from the chaff, for they could say: ``some `gauge transformations' relate distinct physical possibilities while others don't. Thankfully, I, the non-eliminativist, haven't eliminated \textit{any} of them, so I can still tell the two kinds apart!'' This strategy, it is claimed, is not available to Healey. The claim is that, since Healey's eliminativism does not license a distinction between different types of gauge transformations, no restriction to some type of gauge transformation is allowed. In particular, one cannot keep just those transformations that would guarantee invariance of the $\theta_{\text{\tiny YM}}$-term. Therefore Healey would either have to identify  physically distinct states with each other (indeterminacy), or be obliged to set $\theta_{\text{\tiny YM}}$ to zero and thereby fall foul of the fact that allowing for a  non-zero $\theta_{\text{\tiny YM}}$-term is a theoretical requirement. 

\subsection{Our criticism of Dougherty's criticism}\label{sec:JD_intro}

But in fact, no such indeterminacy occurs, since  Dougherty's argument that the $\theta_{\text{\tiny YM}}$-term is only gauge-invariant under gauge transformations that have a particular behaviour at the boundaries is  incorrect. For the $\theta_{\text{\tiny YM}}$-term  is manifestly gauge invariant under the action of {\it all} gauge transformations.

 Nonetheless, \textit{there is a subtle and tempting} reason to erronously assume that the $\theta_{\text{\tiny YM}}$-term is gauge-variant. For, as Dougherty correctly states,  the $\theta_{\text{\tiny YM}}$-term can also be expressed as a pure boundary contribution. And it is well-known that this boundary contribution, which takes the form of a Chern-Simons boundary integral, can acquire different values \textit{even  on vanishing curvature configurations}: the values of such terms can differ by an integer multiple of $2\pi$  (times $\theta$). So, it would be natural to say that these values have some sort of gauge-dependence, i.e. that they  \textit{change}  under ``large gauge transformations''.  
 This  change is the one Dougherty wrongly appeals to in his argument. The mistake is subtle, and lies in the construal of the term ``large gauge transformation''.

 Before we briefly sketch our argument in order to clarify this crucial subtlety, we therefore need to define ``large gauge transformations''.

%


 In practice, the term  ``large gauge transformation'' has been associated with two meanings:\\
 \noindent (\textit{i})  a smooth Lie-group-valued function on  space or spacetime\footnote{The difference between the two is relevant, as we will show later.} that is not connected to the group identity, i.e. not infinitesimally generated through exponentiation; \\
 \noindent (\textit{ii}) in the presence of asymptotic boundaries, it is a gauge transformation which does not asymptote to the identity.
 
In this article, we will exclusively use the term ``large gauge transformation'' in the sense attached to (\textit{i}), i.e. not being connected to the identity.

To make his argument stick, Dougherty must use  transformations that satisfy both  (\textit{i}) and (\textit{ii}) \citep[p. 1]{Dougherty_CP}, i.e. transformations whose pullback to the boundary neither vanishes nor  is connected to the identity: this is because only such transformations would change the value of the boundary Chern-Simons integral which re-expresses the $\theta_\text{\tiny YM}$-term.
However, the combination of (\textit{i}) and (\textit{ii}), required by Dougherty selects an empty set of functions. This is because there is no smooth\footnote{Twice-differentiable is sufficient for our purposes.} Lie-group valued function over $\bb R^4$ that tends at infinity to a function over $\pp \bb R^4 \cong S^3$ that is not connected to the identity. This fact is strictly necessary to ensure the mathematical consistency of the equality between the bulk-integral defining the $\theta_\text{\tiny YM}$-term (which is manifestly gauge invariant under all gauge transformations) and its expression in terms of Chern-Simons boundary integral (which is not invariant under large-gauge transformations over $S^3$). The goal of the following sections is to explain these facts and their basic consequences in good detail. 
 
 Here, we briefly sketch with equations  an abstract argument showing that the necessary transformations cannot be  smoothly extended into the bulk (all notation will be explained later). For now we consider the simplest possible case:\footnote{We will deal with the general case in the following sections.} that of a gauge potential $A$ that is pure gauge on  a 4-disk $D^4$. Thus, $A=g^{-1}\d g$  for some $g: D^4\to G$, and  its associated curvature vanishes, i.e. $F(A)=F(g^{-1}\d g) =0$, so that the $\theta_{\text{\tiny YM}}$-term,  defined as $ \frac{1}{8\pi^2} \int_{ D^4 } \tr( F\wedge F)$, manifestly vanishes---in all gauges.  Thus,
\be\label{eq:intro}
0 = \frac{1}{8\pi^2} \int_{ D^4} \tr( F\wedge F) = \frac{1}{24\pi^2} \oint_{ \pp D^4 = S^3} \tr( g^{-1} \d g \wedge g^{-1} \d g \wedge g^{-1} \d g ) = : \CS_{ S^3}( h^{-1}\d h)
,\ee
where the second equality will be shown in the next section, $\CS_{ S^3}$ is  by definition the Chern-Simons functional (on $S^3$)  with  $ h:S^3 \to G$  set to $h=g_{|S^3}$. 

The puzzle arises thus: it is a mathematical fact that certain $h$'s yield a \textit{non-vanishing} $\CS_{S^3}(h^{-1}\d h)$,  so how could  the above equation avoid mathematical inconsistency? Now, the $h$'s that yield these different values are ``homotopically'' different: they cannot be smoothly deformed into each other, and are thus said to differ by a ``large'' transformation. The answer to our question then is that, crucially,  large transformations that relate different $h$'s  of this kind {\it cannot} be extended into the $D^4 $ bulk smoothly and therefore {\it cannot} define ``gauge transformations" of the bulk configuration $A=0$; there are no such transformations whose restriction to the boundary fits in (\textit{i}) above.
 In other words, the  large \textit{boundary}  transformations required to  yield a non-zero value of the Chern-Simons functional are \textit{not} of the form $h=g_{|S^3}$  for a smooth $g:D^4\to G$. That is: they do not come from  \textit{bulk} gauge  transformations  of any kind---which, as we know, leave the value of the $\theta_{\text{\tiny YM}}$-term invariant.   
 
 Homotopically different $h$'s on the right hand side of \eqref{eq:intro}  represent physically different configurations also in the bulk, and indeed \textit{must} be accompanied by  different curvatures in the bulk. In due course, we will prove all of these statements,  thus avoiding a mathematical contradiction: the gauge-invariance properties of the $\theta_{\text{\tiny YM}}$-term {\it cannot} depend on the way we decide to write it, viz. as a bulk or as a boundary term.

\subsection{Prospectus}
This paper will proceed as follows.
In section \ref{sec:intro0}, we will give a brief introduction to fibre bundles. We start in section \ref{sec:intro} by describing fibre bundles  as the mathematical structure underpinning gauge theories. They formalize the notion that certain properties that are taken as, in a certain sense,  ``intrinsic,'' such as ``being a proton,'' are in fact relational. But these relations can have topological, i.e. global features. Some of these features are embodied by Chern classes, which we briefly review in section \ref{sec:theta}. There, we will recall what these classes have to do with the $\theta_{\text{\tiny YM}}$ term in QCD, and discuss their gauge and topological invariance. In the following subsection \ref{sec:WZ}, we finally bring in the ``large gauge transformations''  that underpin Dougherty's argument and show in particular that they have nothing to do with  gauge-transformations: they are quantities that encode the topological properties of the underlying bundle, and are not  related to choices of gauge. Such topological properties are represented by the particular gluing, or relations, between topologically trivial charts; and the winding numbers encode this `gluing' information. 

These conclusions are valid for manifolds without boundary. In section \ref{sec:bdary} we describe how the previous conclusions can be extended to the context of manifolds with boundaries. Here it is important to distinguish the Euclidean signature setting from the Lorentzian one. In the former case, in section \ref{sec:Euc}, we can complete asymptotic boundaries and fall back on the results for the unbounded manifolds. In the latter, of section \ref{sec:Lor}, we get two disconnected boundaries, and thus  (assuming the fields behave nicely at space-like infinity) the $\theta_{\text{\tiny YM}}$ topological invariant becomes a difference of two Chern-Simons terms, or of two winding numbers. Nonetheless, the conclusions about their invariance remains, but now it applies to the difference of winding numbers. In Section  \ref{sec:conclusions}
we conclude: section \ref{sec:summary} summarizes the main points made in the paper. Finally,  in section \ref{sec:phil}, we briefly smoke a peace-pipe, by giving  a criticism  of our own of eliminativism. This criticism \textit{does} take into account the role of the $\theta_{\text{\tiny YM}}$-term---but not its properties under gauge transformation, which, \textit{pace} Dougherty,   are compatible with eliminativism.

\subsection{Intemezzo: a cautionary remark\label{sec:cautionary}}
 
 Since this article is an answer to  \citet{Dougherty_CP}, we follow here the same,  intrinsically semiclassical, but standard, account of chiral symmetry breaking, cf. e.g. \citep{WeinbergQFT2}. 
 
 However, as we ackowledge in this short intermezzo, a fully non-perturbative account also exists \citep[Ch. 3]{Strocchi_SB}. In the field-theoretic path integral, continuous configurations (which are needed to make sense of the ``topology of the bundle''---see later---as an explanatory device) are of measure zero, and thus  a non-perturbative account of the chiral symmetry breaking mechanism which does {\it not} rely on topological features {\it  of the field configurations} is more satisfying (if not necessary).  In this non-perturbative account, it is rather the topology of the gauge group that plays a crucial role.

Indeed, in the perturbative account of \citet[Ch. 3]{Strocchi_SB}, despite its anomalous implementation, chiral symmetry is not ``explicitly broken'', but rather gives rise to  what one could roughly characterize as a ``meta-symmetry'' between non-communicating  ($\theta$-)sectors of the theory. These sectors are labeled by  their transformation properties under central elements of the algebra of observables  which  correspond to the  equivalence classes of gauge transformations which are not connected to the identity modulo the ones that are connected to the identity (here we refer to the residual time-independent gauge symmetries not fixed by the choice of temporal gauge). The technical, but crucial, ingredient entering this account is the non-weakly-continuous nature of the representation of the symmetries on the Hilbert space.
 
 Although one immediately sees that this non-perturbative account appears quite explicitly incompatible with any naive notion of eliminativism, we will pursue neither a deeper  analysis of its philosophical underpinnings\footnote{ Cf. \citep{Strocchi_phil}.}  nor a clarification of its relationship with the (semi)classical approach;  both are extremely interesting tasks but lie well beyond the scope of this article (but see the comments in sections \ref{sec:thetachern} and \ref{sec:phil}).


\section{Topological invariants and  fiber bundles}\label{sec:intro0}
In this section, we will introduce fiber bundles and their topology, and proceed to assess gauge-invariance of the $\theta_{\text{\tiny YM}}$-term for closed manifolds in several different ways. In section \ref{sec:intro} we introduce the basic machinery: the connection-form (and its relational interpretation), and the relation between charts, gauge transformations and transition functions. In section \ref{sec:theta}, we introduce the $\theta_{\text{\tiny YM}}$-term---also known as the Chern-number. Seen as a bulk, i.e. spacetime, integral, we show both gauge and topological invariance of the term. In section \ref{sec:WZ} we relate this invariant to the appearance of `large' transformations: they appear as Wess-Zumino integrals related to transition functions between charts. We also show that gauge transformations on a 4-dimensional disk-region cannot have non-trivial winding number at its boundary. This is entirely compatible and required by our considerations in this paper.

\subsection{A brief introduction to fibre bundles}\label{sec:intro}
The modern mathematical formalism of gauge theories relies on the theory of principal  (and associated) fibre bundles. We will not give a comprehensive account here (cf. \citep{kobayashivol1}), but only introduce the necessary ideas and objects.

  Given an $n$-dimensional manifold $M$, thought of as representing spacetime (we will not need any of the metric structure of spacetime however), a standard example of a principal fibre bundle with structure group $GL(n)$ taking $M$ as the base space, is the space of linear frames over $M$. The ``fibre'' over each point of the base space $M$ consists in all of the linear frames of the tangent space there. In this example, there is no ``zero'' or identity element on each fibre: each point is just a linear frame basis for the tangent space. But there is a one-to-one map between the group $GL(n)$ and the fibre: we can use the group to go from any frame to any other.  

The main idea underlying the physical significance of the internal space in a fibre bundle is  perhaps best  summarized in the original paper by  \citet{YangMills}: 
\begin{quote} The conservation of isotopic spin is identical with the requirement of invariance of all interactions under
isotopic spin rotation. This means that when electromagnetic interactions can be neglected, as we shall hereafter assume to be the case, the orientation of the
isotopic spin is of no physical significance. The differentiation between a neutron and a proton is then a
purely arbitrary process. As usually conceived, however,
this arbitrariness is subject to the following limitation:
once one chooses what to call a proton, what a neutron,
at one space-time point, one is then not free to make any
choices at other space-time points. \end{quote}
That  is, what is a proton and what is a neutron at a given point is essentially a \textit{relational} property. 

The limitations on how to identify ``a proton'' at two different points of spacetime are imposed by a connection-form: another structure on the bundle. 
That is, a connection-form $\omega$ allows us to define which points of neighbouring fibres can be taken as equivalent to an arbitrary starting-off point in an initial fibre. 
In the example of linear frames, it gives us a notion of  ``parallel transport'' of the basis as we go from an initial choice over one point of $M$, to a neighbouring one. Curvature then acquires meaning as non-holonomicity: start-off from the same point and follow such an identification of bundle points along different paths in the bundle, lifted from a path in $M$, interpolating between initial and final points. Even if the initial point on the bundle and the initial and final points on $M$ agree,  the final  points identified on the bundle may still differ. It is this disagreement that usually carries physical consequences.

 There are two conditions that such a connection-form must satisfy. First, the parallel transport to a neighbouring fibre should commute with the group action; i.e. there is a sense in which it doesn't really depend on what we choose as the starting-off basis.  Equivalently, there is an equivariance property that $\omega$ must satisfy. Secondly, there must be exactly one choice of parallel transported frame per direction of $M$. All the relevant properties of gauge transformations can be derived from these two (cf. footnote \ref{ftnt:omega}). 
 
 We are now going to formalize this intuitive description.

\paragraph*{Principal fibre bundles}

A principal fibre bundle is a smooth manifold $P$ that admits a smooth action of a  (path-connected, semisimple) Lie group, $G$, i.e.  $G\times P\rightarrow P$ with $(g,p)\mapsto g\cdot p$ for some action $\cdot$ and such that for each $p\in{P}$, the isotropy group is the identity (i.e. $G_p:=\{g\in{G} ~|~ g\cdot p=p\}=\{e\}$). 
Naturally, we construct a projection  $\pi:P\rightarrow{M}$, given $p\sim{q}\Leftrightarrow{p=g\cdot{q}}$ for some $g\in{G}$. So the base space $M$ is the orbit space of $P$, $M=P/G$, with the quotient topology, i.e.: characterized by an open and continuous $\pi$. By definition, $G$ acts transitively on each fibre. 

Locally over $M$,  it must be possible to choose a smooth embedding of the group identity into  the fibres. That is, for $U\subset M$, there is a map $\sigma: U\rightarrow P$ such that $P$ is locally of the form $U\times G$, $U\subset {M}$, i.e. there is an isomorphism $U\times G\to \pi^{-1}(U)$ given by $(x, g)\mapsto g\cdot \sigma(x)$.\footnote{Given $p$, the inverse map is a bit more complicated because we must  find $g'$ such that $g'\cdot p=\sigma(x)$, for some $x$. It will depend on the form of $\sigma$. }  The maps $\sigma$ are called {\it local sections} of $P$.

On $P$, we consider an Ehresmann connection $\omega$, which is a 1-form on $P$ valued in the Lie algebra $\mathfrak{g}$ that satisfies appropriate compatibility properties with respect to the fibre structure and the group action of $G$ on $P$.\footnote{Given an element of the Lie-algebra $\mathfrak{g}$, we define the vertical space $V_p$ at a point $p\in P$, as the linear span of vectors of the form $v_{\xi}(p):=\frac{d}{dt}{}_{|t=0}(\exp(t\xi)\cdot p)$ for $\xi\in \mathfrak{g}$. And then the conditions on $\omega$ are:
\[
\omega(v_\xi)=\xi
\qquad\text{and}\qquad
g^*\omega=g^{-1}\omega g,
\]
  where $g^*\omega_p(v)=\omega_{g\cdot p}(g_* v)$ where $g_*$ is the push-forward of the tangent space for the map $g:P\rightarrow P$. A choice of connection is equivalent to a choice of covariant `horizontal' complement to the vertical space, i.e. $H_p\oplus V_p=T_pP$, with $H$ compatible with the group action.  \label{ftnt:omega}}
   This connection allows us to locally define ``horizontal complements'' to the fibres in $P$ (see footnote \ref{ftnt:omega}). Through such complements one can horizontally lift paths $\gamma$ in $M$ to $P$. These horizontally lifted paths are commonly referred to as ``parallel transports" in $P$ along $\gamma$ with respect to (horizontality as defined by) $\omega$.
   As you go around a closed curve  in $M$, parallel transport on $P$ may land you at a different point over the initial fibre from which you started: e.g. assuming you started from $p$, you may end at $p'=g\cdot p$.   The relation between $p$ and $p'$ (i.e.  $g$) is the called the holonomy of $\omega$ along the closed path $\gamma$. Its infinitesimal analogue is the curvature of $\omega$, 
   \be \Omega=\d_{\text{\tiny{P}}} \omega+\omega\wedge_{\text{\tiny{P}}} \omega,
   \ee 
   where $\d_{\text{\tiny{P}}}$ is here the exterior derivative on the smooth manifold $P$, and $\wedge_{\text{\tiny{P}}}$ is the exterior product on $\Lambda(P)$ (it gives anti-symmetrized tensor products of differential forms).

  \paragraph{ Gauge transformations v. Transition functions}
Given  local sections $\sigma_\alpha$ on each chart $U_\alpha$, i.e. maps $\sigma:U_\alpha \to P$ such that $\pi\circ\sigma_\alpha = \mathrm{id}$, we define by pullback $A_\alpha=\sigma_\alpha^*\omega \in \Lambda^1(U_\alpha, \mathfrak{g})$ (here $\alpha$ is a chart index, not a spacetime one).  Since the differential and the pullback operation ``commute'', we  also have:  
 \be\label{eq:curv} F_\alpha:=\sigma_\alpha^*\Omega=\d A_\alpha+A_\alpha\wedge A_\alpha
   \ee 
where now  $\d$ and $\wedge$ are the familiar exterior derivative and products in  $\Lambda(M)$. 

 Notice that contrary to $\omega$ and $\Omega$, the $A_\alpha$'s and $F_\alpha$ are defined over  charts of the spacetime $M$, rather than the bundle $P$. The price to pay is the introduction of: (a) an (arbitrary) choice of section, and (b)---since global sections might not exist in general---of an atlas of charts over $M$ and a corresponding set of $A_\alpha$'s.

In other words, although $\omega$ is globally defined on $P$, the $A_\alpha$'s are only defined on the respective charts $U_\alpha$ of ${M}$ through the choice of a local section $\sigma_\alpha$. At fixed $\omega$, and on a given chart $U_\alpha$, different choices of section give $A_\alpha$'s related by a gauge transformation.  The demand of gauge invariance reflects the arbitrary nature of the choice of section. We will come to this in a moment; first we need to worry about how to patch the charts together.

Given an atlas of charts $U_\alpha\subset M$, this patching requires us to consider transition functions
which relate the $A_\alpha$'s to each other on the overlaps $U_{\alpha\beta}=U_\alpha\cap U_\beta$: 
\be\label{eq:transition}
\text{on $U_{\alpha\beta}$: \quad} A_\beta = \t_{\alpha\beta}^{-1} A_\alpha \t_{\alpha\beta} + \t_{\alpha\beta}^{-1} \d \t_{\alpha\beta},
\ee
where 
\be
\t_{\alpha\beta}\equiv \t_{\beta\alpha}^{-1} : U_\alpha\cap U_\beta\to G.
\ee

These transformation properties translate between choices of local sections across overlapping charts, and must satisfy the cocycle conditions (compatibility over threefold overlaps $U_{\alpha\beta\gamma}=U_\alpha\cap U_\beta \cap U_\gamma$):
\be
\text{on $U_{\alpha\beta\gamma}$: \quad}\t_{\gamma\beta}\t_{\beta\alpha} = \t_{\gamma\alpha}.
\label{eq:cocycle}
\ee
Transition functions look similar to  gauge transformations, and indeed act very similarly on the gauge potentials. These similarities reflect the fact that, on the overlap $U_{\alpha\beta}$, both $A_\alpha$ and $A_\beta$ descend from the same $\omega$ through different choice of sections---and, as we will now discuss, the role of gauge transformations is precisely to translate between different choices of sections.

Gauge transformations  (i.e. changes of local sections) are   encoded in maps\footnote{The set of all $g_\alpha$'s on a given $U_\alpha$ defines $\G_\alpha:=\{g_\alpha(x)\}$, which inherits from $G$ the structure of an (infinite-dimensional) Lie-group, by pointwise extension of the group multiplication of $G$ over $U_\alpha$.} 
\be\label{eq:gt}
g_\alpha : U_\alpha \to G
\ee 
that act on the respective $A_\alpha$ {\it and}  $\t_{\alpha\beta}$'s as follows:
\be
\begin{dcases}
A_\alpha \stackrel{g}{\mapsto} A_\alpha^g = g_\alpha^{-1} A_\alpha g_\alpha + g_\alpha^{-1} \d g_\alpha & \text{on $U_\alpha$}\\
\t_{\beta\alpha} \stackrel{g}{\mapsto}  \t_{\beta\alpha}^{g} = g_\beta^{-1}\t_{\beta\alpha}g_\alpha & \text{on $U_{\alpha\beta}$}
\end{dcases}
\label{eq:gaugetransf}
\ee
from which one derives using \eqref{eq:curv}:
\be
F_\alpha \stackrel{g}{\mapsto} F_\alpha^g = g_\alpha^{-1} F_\alpha g_\alpha  \quad \text{on $U_\alpha$}.
\label{eq:gaugeF}
\ee
 Notice that \textit {both the connection and the transition function} transform under the action of a gauge transformation $g_\alpha$. Thus, under a gauge transformation on $U_\alpha$, equation \eqref{eq:transition} describing the {\it relation} between $A_\beta$ and $A_\alpha$, is left invariant. {\it This is the basic reason why the transition functions collectively encode the global properties of the bundle $P$ while the gauge transformations are simple redundancies.}

 Besides the fact that gauge transformations act on transition functions and not vice versa,  another crucial distinction between gauge transformations and transition functions, that underlies their different roles, is that the domain of  the gauge transformations $g_\alpha$'s is the {\it whole} $U_\alpha$,  whereas that of $\t_{\alpha\beta}$ is  a subset of $U_\alpha$  (viz. its overlap with $U_\beta$). 

We reiterate that the introduction of transition functions is generally necessary because, global sections do not exist unless the bundle is trivial, i.e. unless $P= M \times G$ {\it globally} not just locally. In the trivial case, and only in the trivial case, all transition functions can be trivialized to be the identity, i.e. $\t_{\beta\alpha} = g_\beta g_\alpha^{-1}$ for some choices of $g_\alpha$'s.  Only {\it then}, equation \eqref{eq:transition} is trivialized and the collection of $A_\alpha$'s yields a global gauge potential 1-form $A$.

\paragraph{Summary} A gauge field configuration can be defined either:\\
\noindent (1) ``abstractly,'' by providing a bundle $\pi:P\to M$ and an Ehresmann connection $\omega\in\Lambda(P,\mathfrak g)$; or\\
\noindent (2) ``in coordinates,'' by providing an atlas of charts $U_\alpha\subset M$, a set of sections $\sigma_\alpha: U_\alpha \in P $, and compatible\footnote{Compatibility is here understood in the sense of equations \eqref{eq:cocycle}.} transition functions $\t_{\alpha\beta}:U_{\alpha\beta}\to G$ (these three ingredients define $P$), together with a choice of compatible\footnote{Compatibility is here understood in the sense of equations \eqref{eq:transition}.} gauge fields $A_\alpha\in\Lambda^1(U_\alpha,\mathfrak g)$ (this corresponds to the choice of $\omega$). 

The coordinate description is redundant because it requires the introduction of auxiliary choices of sections, $\sigma_\alpha$; different choices are related by ``gauge transformations'' of the $A_\alpha$'s {\it and} of the $\t_{\alpha\beta}$'s. Therefore, gauge invariance requires all physical observables to depend on the choice of $P$ and $\omega$ only.\footnote{Notice that it is possible to change $\omega$ (resp $A_\alpha$) without changing $P$ (resp $\sigma_\alpha$ and $\t_{\alpha\beta}$).}

Crucially, transition functions and gauge transformations play entirely different roles. 
Gauge transformations act on the transition functions, but not vice-versa, and  a gauge transformation's domain of definition  is the {\it whole} chart $U_\alpha$, and not merely the overlaps $U_{\alpha\beta}$ as is the case for the transition functions $\t_{\alpha\beta}$'s. These technical differences  reflect the fact that the $g_{\alpha}$'s and $\t_{\alpha\beta}$'s play conceptually different roles. 
 From the perspective of $P$, the gauge transformations $g_\alpha$'s encode the freedom of choosing a local section $\sigma_\alpha$ (which is necessarily defined on the whole $U_\alpha$). Conversely, the $\t_{\alpha\beta}$ encode---albeit somewhat redundantly---the way in which the charts are glued to one another, and thus the global structure of the bundle $P$.

\subsection{The Chern-number}\label{sec:theta}
For a closed $4$-dimensional manifold ${M}$, that is, ${M}$ is compact and without boundary, the quantity (the notation will be explained in a moment,  for now it is enough to notice that the integrand depends on $A$ and is gauge-invariant)
$$
\Ch[P] :=\int_{{M}} \ch_A
$$
is a topological invariant---{\it not} of ${M}$---but of the {\it fibre bundle $P$ over ${M}$}. A connection-form $\omega$ is defined over $P$ and a collection of local gauge potentials $A_\alpha$ is defined over  an atlas of ${M}$, as above.   Since $\ch_A$ is gauge invariant, the integral can then be obtained through an appropriate partition of unity associated to the atlas. As a topological invariant of $P$, $\Ch[P]$ is not only completely gauge invariant, but also independent of the choice of $\omega$ over $P$. We call $\Ch[P]$  the (second) \textit{Chern-number} of $P$.

If we write our physics in terms of  gauge potentials, and allow them to live in different bundles, e.g. $P$ and $P'$,  then the potentials $A$ and $A'$ might lead to different values of $\Ch[P]$. The question then is: how does $A$ ``know about''  topological properties' of $P$? And how can $\Ch[P]$ depend only on the topology of $P$ and not on the detailed choices going into its computation? This is the content of the Chern-Weil theorem \cite[Ch. 11.1]{Nakahara}, that we  briefly review below.

 From now onwards, we will restrict to $G = \SU(N)$.


First, the Chern-number is computed as follows: 
\be
\label{eq:theta}
\Ch[P]=\int_M \ch_A=\frac{1}{8\pi^2}\int_M \tr (F\wedge F)
\ee
where
\be
\ch_A:= \frac{1}{8\pi^2}\tr( F\wedge F).
\ee 
Of course, $\Ch(P)$ is nothing but the  ``$\theta_{\text{\tiny YM}}$-term,'' or, more specifically: the $\theta_{\text{\tiny YM}}$-term in the QCD Lagrangian can be written using \eqref{eq:theta} as:
 \be
  \mathcal{L}_\theta=  \theta \,\Ch[P]
 \ee
where $\theta$ is just a real-valued coefficient.
The integrand $\ch_A$ defines the \textit{second Chern-class} of the bundle $P$. The second Chern-class is manifestly gauge invariant, given the gauge transformation properties of $F$ \eqref{eq:gaugeF} and the cyclicity of the trace.\footnote{The proof is simple: $\tr(g^{-1}Fg\wedge g^{-1}F g)=\tr(g^{-1}F\wedge F g) =\tr(  Fg \wedge g^{-1}F)=\tr(F\wedge F)$.\label{ftnt:F_proof}} This means that on the overlaps $U_{\alpha\beta}$,  $ \ch_{A_\alpha} = \ch_{A_\beta}$, which is why no chart index appears in the equations above, and why the integral can be performed with no further complications. 

This also immediately tells us that $\Ch[P]$ can at most depend on the choice of $\omega$, and not of gauge (i.e. of sections). We are now ready to review the Chern-Weil theorem, which shows that $\Ch[P]$ is not only gauge invariant but also independent of the choice of $\omega$ on $P$---that is it depends only on the topological properties of $P$. 

A first hint of the `topological' nature of $\Ch[P]$ comes from the observation that it does not change under a small arbitrary variation of $A$ (i.e. the equations of motion of the action $S[A] = \int \ch_A$ are identically satisfied). This follows  immediately from $\delta F=\d_A\delta A$ and the Bianchi identity $\d_A F=0$ where $\d_A:=\d+[A, \cdot]$ is the exterior gauge-covariant derivative (for the adjoint representation). 
But invariance can be proven  also for finite, rather than infinitesimal, changes in connection. Consider two connections $A$ and $A'$, and now define $\gamma := A' - A \in \Lambda^1(M)$ and a one-parameter family of connections $A_s=A+s\gamma$, $s\in(0,1)$, interpolating between $A$ and $A'$ (the space of connections is an affine space). Then, denoting the curvature of $A_s$ as $F_s$, one finds
\be
\ch_{A'}-\ch_{A}\equiv\frac{1}{8\pi^2} \int^1_0 \frac{\d}{\d s}\tr( F_s\wedge F_s)\d s=\frac{1}{4\pi^2}\int_0^1\tr(\d_{A_s} \gamma\wedge F_s)\d s=\frac{1}{4\pi^2} \d \Big(\int_0^1\tr(\gamma\wedge F_s)\d s \Big), 
\ee
Thus the difference $ \ch_{A'}-\ch_{A}$ is an exact differential form and thus vanishes when integrated over a closed manifold.\footnote{For consistency, one should also check that the the 3-form $\int_0^1\tr(\gamma\wedge F_s)\d s$ is well defined, i.e. gauge invariant. That this is the case follows from the fact that the difference $\gamma$ between two connections transforms in the adjoint representation under gauge transformations, just like $F$, and therefore $\tr(\gamma\wedge F_s)$ is point-wise  gauge invariant for all values of $s$. (cf footnote \ref{ftnt:F_proof}).} Since $A$ and $A'$ are arbitrary connections, it follows that $\int_M \ch_A$ over a closed manifold $P$ does not depend on the choice of connection, i.e. that it is a topological invariant. 

\paragraph{Summary} The gauge invariance of $\ch_A$ tells us that $\Ch[P]$ depends at most on $\omega$, and the Chern-Weil theorem tells us that $\Ch[P]$ does not depend on $A$ (and therefore on $\omega$) at all. Therefore, $\Ch[P]$ can only reflect a (topological) property of the bundle $P$ on which the connection is defined. 
A  nontrivial, and extremely deep, fact is that the second Chern number of $P$ is always an integer
\be
\Ch[P] \in \bb Z.
\ee

%

We conclude this section with a simple remark.
The discussion above clearly shows that the Chern number  \eqref{eq:theta} (and thus the $\theta_{\text{\tiny YM}}$-term) is gauge-invariant under all possible gauge transformations. And, just to be clear, this even holds at the level of the integrands:
\be
\ch_{A^g}=\ch_A \qquad \forall g=g(x)
\label{eq:ch_inv}
\ee
This fact simply follows from the transformation properties of $F$ \eqref{eq:gaugeF} and the (graded) cyclicity of the trace (for $\lambda, \eta$ as p and q-forms, respectively)
\be
\label{eq:cyclic}
\tr (\lambda\wedge \eta)= (-1)^{pq}\tr (\eta\wedge \lambda).
\ee
Therefore any non-gauge invariance of the $\theta_{\text{\tiny YM}}$-term is vetoed by this simple demonstration.

\subsection{Transition functions and large gauge transformations}\label{sec:WZ}

As we have just witnessed, the Chern-number and the so-called $\theta_{\text{\tiny YM}}$-term,  \eqref{eq:theta}, is \textit{completely} gauge-invariant. Thus the inevitable question: whence Dougherty's claims? He writes for example that (italic ours)
\citep[p. 7]{Dougherty_CP}\begin{quote}
{\it The Yang- Mills $[\theta\text{-}]$vacuum term is
not preserved by all gauge transformations.} If the eliminative view
of gauge transformations is right, this means that the Yang-Mills
vacuum term is physically meaningless. If gauge transformations
are redundancies then mathematical differences between gauge
equivalent
configurations can't reflect physical differences. So the
value of the Yang-Mills vacuum term can't represent any physical
fact.
\end{quote}
We will now argue that one way Dougherty might have arrived at this conclusion, ignoring the previous simple argument for the gauge invariance of the $\theta_{\text{\tiny YM}}$-term, is through an incatious invocation of boundaries. 

Before we get to boundaries of the entire Universe, in section \ref{sec:bdary}, let us revisit the computation of the Chern-number under a new guise, by breaking up the manifold and therefore introducing internal boundaries.  

First, we recall the Chern density 
$
\ch_A:= \frac{1}{8\pi^2}\tr( F\wedge F)$, and the following crucial relation\footnote{This is easy to show: 
\begin{eqnarray*}
\d\cs_A&=&\d{\tr}(A\wedge \d A+\tfrac23A\wedge A\wedge A)={\tr}(\d A\wedge \d A+2A\wedge A\wedge \d A)\\
&=&{\tr}((\d A+A\wedge A)(\d A+\wedge A\wedge A))=\ch_A
\end{eqnarray*}
where in going from the first to the second line we used \eqref{eq:cyclic} to infer that $\tr(A\wedge A\wedge A \wedge A)\equiv0$.\label{proof}} it has with the \textit{Chern-Simons 3-form} $\cs_A$:\footnote{ The Chern-Simons functional understood as the action for a 3d boundary theory, defines a classical theory of connections that is  invariant only under gauge transformations that are {\it not} large in the sense of (\textit{i}) in Section \ref{sec:JD_intro}.  However, quantum mechanically, the situation can be improved, and the Chern-Simons functional can define a theory which is invariant under {\it all} gauge transformations, provided the coupling constant, i.e. the Chern-Simons ``level'', is chosen to be an integer. This is because under large gauge transformations, the Chern-Simons action changes at most by a multiple of $2\pi$---hence allowing the Feynman's path integral to still be invariant. This peculiarity lies at the root of the fascinating phenomenology of Chern-Simons theory and its quantum-deformed symmetry structure.\label{fnt:qCS} }  
\be
\ch_A = \d \cs_A
\qquad\text{where}\qquad
\cs_A := \frac{1}{8\pi^2} \tr( A \wedge \d A+ \tfrac23 A\wedge A \wedge A) .
\label{eq:fundam}
\ee

The subtlety lurking behind this identity is the fact that the Chern-Simons form is, at least naively, \textit{not} gauge invariant, since:
\be
\cs_{A^g}-\cs_{A} = \wz_g +\frac{1}{16\pi^2} \d \;\tr( \d g g^{-1} \wedge  A) 
\label{eq:lingo}
\ee
where the Wess-Zumino term is just a Chern-Simons form of a pure gauge configuration: 
\be
\label{eq:wz}
\wz_g := \cs_{g^{-1}\d g}=\frac{1}{24\pi^2}\tr(g^{-1}\d g\wedge g^{-1}\d g\wedge g^{-1}\d g).
\ee
In particle physics lingo, equations \eqref{eq:ch_inv}, \eqref{eq:fundam}, and \eqref{eq:lingo} together say that ``while the topological charge [$\ch_A$] is gauge invariant, the topological current [$\cs_A$] is not.'' \citep[p. 31]{CP_Schafer}. 

However, as demanded by mathematical consistency  between the invariance of $\ch$ and its relation to $\cs$ in \eqref{eq:fundam}, both sides of \eqref{eq:lingo} must be closed 3-forms, and therefore   $\wz_g$ is necessarily a closed 3-form, i.e.\footnote{This is a corollary of the fact that $\tr(A\wedge A \wedge A \wedge A)\equiv 0$ (see footnote \ref{proof}), since $\d (g^{-1} \d g) = - g^{-1} \d g \wedge g^{-1} \d g$.} 
\be
\d \wz_g \equiv 0.
\ee
Therefore, the gauge invariance of $\ch_A$ is not affected, even if we write it in terms of the gauge-\textit{variant} functional $\cs$:
\be\label{eq:lingo2}
\ch_{A^g} = \d \cs_{A^g} = \d ( \cs_A + \wz_g + \d\; \frac{1}{16\pi^2}\tr( \d g g^{-1} \wedge  A) ) = \d \cs_A = \ch_{A}. 
\ee
  In particular, taking $A=0$ and integrating this equation on a manifold with boundary, we see that the boundary integral of the Wess-Zumino term associated to a gauge transformation in the bulk {\it necessarily} vanishes. 
   Equation \eqref{eq:lingo2} is a first important check, which we will now corroborate with a different calculation.

This different computation resolves possible confusion having to do with a particular way of expressing $\Ch[P]$. Namely, 
 there is still one manner of computing  $\Ch[P]$  chart by chart, using   \eqref{eq:fundam}, which may confusingly appear gauge-variant. We will now set up the puzzle and then dissolve it.  Instead of dealing with these issues on a very general basis, we specialize our discussion to a more concrete example. 

Consider  the closed manifold $ M = S^4$ covered by 2 charts, isomorphic to 4-dimensional disks, $U_{1}, U_2=D^4$, that overlap on a ``transition belt'' around the equator, $U_{12}=S^3\times[-1,1]$. 

We know that at the interface, by \eqref{eq:transition},  $A_1={A^{\t}_2}$, $\t \equiv \t_{21}$. 
Denoting the domain of the charts that lies below/above the equator, respectively, by $\tilde U_1 = U_1 \setminus (S^3\times [-1,0])$ and $\tilde U_2 = U_2 \setminus (S^3\times[0,1])$ (notice that $\pp \tilde U_1 = - \pp \tilde U_2 = S^3 \times \{0\} \simeq S^3\subset U_{12}$), we have 
\begin{align}
\Ch[P] 
&= \int_{\tilde U_1} \ch_{A_1} +  \int_{\tilde U_2} \ch_{A_2}    = \oint_{\pp \tilde U_1}( \cs_{A_1} - \cs_{A_2})   =  \oint_{\pp \tilde U_1}( \cs_{A^{\t}_2} - \cs_{A_2}) =  \oint_{\pp \tilde U_1} \wz_{\t} \label{eq:intersec}
\end{align}
where we used \eqref{eq:lingo} and \eqref{eq:wz}. 

Thus we see that, setting ${\pp \tilde U_1}\simeq S^3$ and denoting $\WZ_{S^3}(g) = \int_{S^3} \wz_g$, 
\be
\bb Z \ni \Ch[P] = \WZ_{S^3}(\t).
\ee
This equation is of crucial importance for us. We have not used gauge transformations, and yet, something that ``looks like'' a gauge-transformation, namely, a transition function \eqref{eq:transition} has appeared in the computation. Now we will verify that we cannot get change the Wess-Zumino invariant related to $\t$ by applying a gauge transformation. 

First of all, as discussed in section \ref{sec:intro}, $\t$ encodes a topological property of the bundle. It is therefore not to be interpreted as a gauge transformation, but as part of the definition of $P$. 
But things are subtle, because---as we summarized in the last paragraph of section \ref{sec:intro}---$\t$ participates in the definition of $P$ in a way that depends on the choice of gauge, i.e. of sections $\sigma_\alpha$. As a consequence, under a change in the choice of sections, the transition functions transform according to \eqref{eq:gaugetransf}:
\be
\t \mapsto g_{2}^{-1} \t g_1.
\ee
Thus, the question arises: why does the following equality, 
\be
\WZ_{S^3}(\t) = \WZ_{S^3}(g_2^{-1} \t g_1),
\label{eq:wz=wz}
\ee
hold?

From a \textit{strictly} three-dimensional, or boundary,  perspective there is no reason why this should be the case. In particular, we could always choose $g_1 = e $ (the identity of $G$) and $g_2$  such that  $(g_2)_{|U_{12}}= \t$, thus apparently trivializing the value of $\WZ_{S^3}$.
However, {\it once we take into account the   whole domain of definition of the $g_\alpha$'s}, which extends into the four-dimensional bulk of the two hemispheres, the above choice might simply be \textit{unavailable}. That is, if $\t : S^3 \to G$ is large according to sense (\textit{i}) in Section \ref{sec:JD_intro}---not connected to the identity---there is \textit{no} smooth extension of it that goes from the belt overlap $U_{12}=S^3$ to the chart domain $U_2 = D^4$. An extension would  necessarily have to  ``break'' somewhere inside $U_2$. Only for $\t$'s connected to the identity will there be a smooth $g_2$  such that  $(g_2)_{|U_{12}}= \t$.  

We can easily perform a proof by contradiction (\textit{reductio}). For suppose it was possible to smoothly extend such $g_\alpha$'s into the interior of their charts,  then, following a radial evolution in the disk $U_2=D_4$, we would find a $g(x,r)$ such that $g(x,r=1) = \t(x)$ and $ \lim_{r\to 0} g(r, x)= g_o$ for all $x\in S^3$,  where $g_o$ is some fixed element of $G$. 
But  exploiting this radial parametrization we can define a 1-parameter family of gauge transformations $\{ h_r(x):S^3 \to G \, | \, h_r(x) = g(r,x) \}_{r\in [0,1]}$, defined \textit{at} the intersection $S^3$, such that $\WZ(h_{r=0}=g_o)=0$ and $\WZ(h_{r=1}=\t)\neq 0$. But this cannot be right: $\WZ(h_r)\in\bb Z$, and since one cannot continuously jump between discrete values, $\WZ$ has to  be constant on path-connected components of its domain.  Let us  prove this explicitly  (by adding a differentiability assumption): denoting $h_r(x) = g(r, x)$ and $\xi_r = \frac{\d h_r}{\d r}h_r^{-1} $, we have, for an arbitrary $r=r_o$,
\be
\frac{\d}{\d r}\WZ_{S^3}(h_r) {}_{|r=r_o}= \oint_{S^3} \frac{\d}{\d r}\wz_{h_r}{}_{|r=r_o} = \frac{1}{24\pi^2}\oint_{S^3}  \d\; \tr(  \d \xi_{r_o}  \wedge h_{r_o}^{-1} \d h_{r_o}) = 0
\ee
where the second equality 
follows from \eqref{eq:wz}.

 In more pictorial terms, $\WZ_{S^3}(h)$ computes a ``winding number'' of the map $h : S^3 \to G$; this is a topological quantity that cannot be undone by a smooth deformation of $h$. However, any \textit{smooth map} $g_\alpha(x,r)$ from the 4-disk $D^4$ into $G$---a gauge transformation according to (\textit{i})---\footnote{It is clear that transformation which are not smooth to some degree are not allowed. Here we only need them to be $C^2$.} automatically provides through ``radial evolution'' a homotopy of maps $h_r(x) = g_{\alpha}(r,x):S^3\to G$ between a {\it constant} function $h_{r=0}(x) = \lim_{r\to 0} g_{\alpha}(r,x) = g_o$ (at the central point)  and its boundary value $h_{r=1}(x) = g_{\alpha}(r=1,x)$. 
 
 It follows that the boundary value of a \textit{bulk gauge transformation} $g_\alpha$ must have trivial winding number as a map from $\pp U_\alpha \to G$, i.e. $\WZ_{S^3}(g_\alpha{}_{|\pp U_\alpha}) \equiv 0$. That is,  the boundary value of \textit{any} gauge transformation $g_{\alpha}(x, r=1)$ on such charts  must be connected to the identity.

From this, it readily follows that $\t$ and $g_2^{-1}\t g_1$ are in the same homotopy class as maps from $S^3$ into $G$, and therefore have the same winding number, as per equation \eqref{eq:wz=wz}.

Therefore, we conclude that in the simple case analyzed here the second Chern number of the bundle $\pi:P\to S^4$ is fully encoded into the winding number of the ``equatorial'' transition function $\t : S^3 \to G$. This winding number is an intrinsic property of $\t$ that cannot be changed by any gauge transformation. 


So far we have discussed bundles on manifolds without boundaries. But to satisfactorily vanquish all doubts about gauge-invariance, we should also guarantee that it emerges when the $\theta_{\text{\tiny YM}}$-term is expressed  not at intersections, but at boundaries. This is only possible when the curvature vanishes at the boundary; e.g. asymptotically. We now turn to this.

\section{Manifolds with boundaries}\label{sec:bdary}

In the first section, \ref{sec:Euc}, we will examine Chern classes within a single bounded, Euclidean manifold and its relation to the Chern-Simons and Wess-Zumino functionals.   In section \ref{sec:Lor} we briefly examine the Lorentzian case, with two boundaries, one asymptotic past Cauchy surface and one asymptotic future one (as most of the literature;  see e.g. \citet[p.454-455]{WeinbergQFT2}) we neglect spatial boundary terms at infinity  (on which $A$ is supposed to vanish). The Chern class then gives a difference of past and future Chern-Simons terms, (naively) representing a transition between different vacua of the theory. In section \ref{sec:thetachern}, we briefly discern the meaning of non-trivial bundle topology viz. the meaning of individual winding numbers.

\subsection{In Euclidean signature.}\label{sec:Euc}
Setting aside an exhaustive treatment of fibre bundles over manifolds with boundaries, which goes beyond the scope of this article, we will content ourselves with discussing what happens first for  $M\cong D^4$ with a boundary $S^3$, and then for $M \cong \bb R^4$ complemented with its asymptotic boundary $B^3_\infty \cong S^3$.  

 First, we recall that gauge transformations on $D^4$  induces gauge transformations on $\pp D^4=S^3$ that are necessarily connected to the identity (as 3d objects). Armed with this fact, we can already see why our conclusions of gauge-invariance will hold in the bounded case: even if different enough $A$'s  give different Chern-numbers (since they may yield different Chern-Simons terms at the boundary, according to \eqref{eq:fundam}), such $A$'s would not be related by a gauge transformation, as guaranteed by   equation \eqref{eq:lingo2}. This proof was easy, but it doesn't yet get to the bottom of the puzzle, which we can only articulate when expressing such integrals in terms of winding numbers, i.e. Wess-Zumino functionals. And for that, we need boundary conditions guaranteeing that the curvature vanishes,\footnote{Note that, for internal boundaries, i.e. for the intersection between charts, we can express the integrals in terms of Wess-Zumino integrals, as in \eqref{eq:fundam}, because it depends on the difference between two Chern-Simons functionals, and smoothness guarantees that this difference can be expressed purely in terms of the transition functions; i.e. Lie-group valued functions. } which we can treat jointly with the asymptotic case.

Topologically, the space $M \cong \bb R^4$ is just\footnote{Following Penrose (cf. \citet[Sec. 5]{hawking1975large}), the physically meaningful way to complement $\bb R^4$ with a boundary depends on its metric (which so far has played no role whatsoever in our considerations).  The choice followed here corresponds to the Euclidean 4-dimensional world, rather than a Minkowskian one (which requires the introduction of five different typologies of asymptotic boundaries: future and past time-like infinity, future and past null infinity, and spacial inifinity). However, ignoring this complication  might be justified since the metric one picks on $\bb R^4$ does not matter for the computation of the $\theta_{\text{\tiny YM}}$-term.  Indeed, the computation in \citep[Sec. 23.6]{WeinbergQFT2} also disregards these subtleties. However, we personally find this argument not completely satisfactory. For now, we leave  this subtle point aside. \label{fnt:Penrose}} a 4-disk, and we denote it $\bb R^4_\infty \cong D^4$  to emphasize the addition of a sphere at infinity, $\pp \bb R^4_\infty = B^3_\infty \cong S^3$. The simple remark that $D^4$ constituted one of two hemispheres in the previous discussion will become useful later.

The gain is that, now, a single chart  covers the whole space; the loss is that this raises a puzzle: without any need for a transition function, what is left of the previous arguments we applied for the $\WZ$ term?

As standard, we start by requiring that the field strength vanishes sufficiently fast at infinity to render the Yang-Mills action, supplemented by the $\theta_{\text{\tiny YM}}$ term, finite. This implies in particular that the gauge potential must approach a curvature-free  configuration at infinity:
\be
A \xrightarrow{x\to\infty} h^{-1} \d h \quad\text{for some}\quad h : B^3_\infty\cong S^3 \to G.
\label{eq:as_bc}
\ee
 Note that this $h$ need not be seen as a gauge transformation---vanishing curvature guarantees \eqref{eq:as_bc}---and thus a characterization as ``pure gauge'' can be misleading.  For such an $h$ may still `wind around' the boundary, in which case $A$ cannot be of the form $A=g \d g^{-1}$ throughout the region. That is, an $A$ that has non-trivial winding number at the boundary must have curvature in the bulk.\footnote{The proof follows the one showing a gauge transformation can only have a trivial winding number, in the previous section.}

For such an $A$, from \eqref{eq:fundam} and \eqref{eq:wz} one has:
\be
\int_{\bb R^4_\infty} \ch_A= \int_{B^3_\infty}\wz_{B^3_\infty} (h).
\ee
(we avoid the Chern-number notation, $\Ch$,  because we do not have a closed base manifold, this preferrence will be maintained in what follows). 
 Again, we know that \textit{no} gauge transformation---which by definition must be extendible into $\bb R^4_\infty$---can be large at the boundary, nor can it change the local value of $\ch_A$, and therefore none can change the value of {\it either} of the integrals above. This quantity {\it is} therefore fully gauge invariant, just as the left-hand side shows manifestly. 

Intriguingly, even in this, single-boundary case, the Wess-Zumino invariant is still an integer! Of course, had we computed the quantity  $\int\ch_A$ with arbitrary boundary conditions, we can get any (gauge-invariant) quantity, depending on the boundary conditions.  $\WZ_{B^3_\infty}(h)$ is valued in the integers because of the asymptotic conditions required on the gauge potentials, which are necessary for the integral to converge. As before, this integer counts how many times the boundary map $h:S^3 \to G$ winds around the group.

A deeper reason why this integral still yields an integer is that, due to the boundary conditions, it can be recast as an integral over a closed manifold, as before. That is, in the Euclidean case being studied here, we can connect the above computations with the previous ones performed for the closed manifold case, at the end of Section \ref{sec:WZ}. It turns out that given the asymptotic boundary conditions \eqref{eq:as_bc}, there is a ``minimal'' way to extend the bundle over $M= \bb R^4_\infty\cong D^4$ to a bundle $\bar P$ over a closed manifold $\bar M \cong S^4$ (where we denote the closure by an overbar). Then, with this extension,
\be
\Ch[\bar P] = \int_{\bb R^4_\infty} \ch_A.
\ee

To understand $\bar P$, it is enough to observe that the asymptotic boundary conditions \eqref{eq:as_bc} are just the minimal\footnote{Here we are ignoring subtleties related to rapidity of the fall-offs at infinity and smoothness in the compactified manifold.} requirements to be able to compactify $\bb R^4$ to $S^4$. 
If the field strength vanishes at infinity rapidly enough, we can compactify $\bb R^4$ to $S^4$ by simply adding {\it one}\footnote{As opposed to a three-sphere.} point at infinity---the North Pole in the stereographic projection of $S^4$---and declaring that at this point $F=0$---the only value it can assume by continuity.  This compactification will take us back to our  previously covered example. 


\subsection{In  Lorentzian signature}\label{sec:Lor}
 But there is still one remaining piece of the puzzle. Much of what we have done is based  on an Euclidean-signature intuition for the manifold $\bb R^4_\infty$: the $\theta_{\text{\tiny YM}}$-term measures the topology of a canonically defined bundle on $\bar P \to S^4$ and $\WZ_{S^3_\infty}(h)$ measures the winding number of the asymptotic field configuration around the 3-sphere at infinity. Thinking about the Lorentzian case opens new perspectives.
 
To think about the manifold with Lorentzian signature,  we can imagine squishing the boundary at infinity  $B^3_\infty \sim S^3$ from opposite sides, making it look more and more like a `thin lens'. This effectively  separates the boundary into three components: a past and a future Cauchy surface, $\Sigma_{\pm}$,  and a ``celestial sphere'' $S^2_\infty$ at spatial infinity.\footnote{See however footnote \ref{fnt:Penrose}.}
Each  Cauchy surface supports some (asymptotic) gauge-potential configuration that encodes a classical state of the theory.  In our case, these states have   half of  their support on the northern (southern) hemisphere of  $S^3_\infty$ corresponding to the asymptotic past (future, respectively) Cauchy surfaces. 

 It is easy to find configurations that are curvature-free at asymptotic past and future infinities, $\Sigma_{\pm\infty}$. For the same reason as in the previous case,\footnote{ Together with assumptions about the field behaviour at spatial infinity, see e.g. \citet[Ch. 23.5 p. 454-455]{WeinbergQFT2}.} asymptotic conditions guarantee that the Chern-Simons terms are numbers, $n_\pm$. And  due to the fixed orientation of these surfaces,  the Chern class gives a difference between these numbers, i.e. $\int\ch_A=n_+-n_-$. 

Therefore, in a similar fashion to what we did throughout the paper, we can reconcile the fact that \textit{curvature-free} boundary states  $h$  \eqref{eq:as_bc}  can encode the physical, i.e. gauge-invariant, value of the $\theta_{\text{\tiny YM}}$-term---\textit{which only depends on the curvature}.  

To summarize some of these results from different contexts: while it is true that only the curvatures figure in the argument of  $\int\ch_A$,   this term is only related to  Chern-Simons terms \textit{on the boundaries of the manifold} (cf. \eqref{eq:fundam}), and these latter terms do \textit{not} depend on the curvature. For closed unbounded manifolds, winding numbers appear as differences of Chern-Simons terms at transition patches; for Euclidean bounded manifolds, the boundary is connected and we obtain a single winding-number (that cannot be changed by gauge transformations that properly extend into the bulk); but here, since the configurations are ``pure gauge'' at disconnected boundaries, we  extract winding numbers from each connected boundary Chern-Simons term. The $\theta_{\text{\tiny YM}}$-term, $\int \ch_A$,  will thus be related to a difference of  winding numbers  due to the inward/outward orientation of the two Cauchy slices with respect to the 4-dimensional bulk. 

 But,  as emphasized after equation \eqref{eq:as_bc}, curvature-free vacuum states with different nontrivial winding numbers,\footnote{ Extra conditions at $S_2^\infty$ may be needed to have well defined winding numbers on the past and future Cauchy surfaces independently. We will ignore this issue,  since we can resolve the puzzle without it.} although perfectly admissible, must include curvature in the bulk. 
This means that, although the individual boundary winding numbers associated to each boundary are not distinguishable by curvature invariants, transitions between them are. 
And this is because, crucially, the transition between different curvature-free boundary {\it states} with non-trivial winding numbers can never proceed through curvature-free {\it histories}.\footnote{This follows from the same arguments exposed below equation \eqref{eq:wz=wz}.} Within the bulk of  spacetime, one has to go through non-vanishing values of $F$ that contribute to $ \ch_A$, and   values which are uncontroversially encoded in the holonomies.

\subsection{ Non-trivial bundle topology and the $\theta$-vacuum}\label{sec:thetachern}

 The quantity $\int \ch_A$  itself is  computable even from an eliminativist perspective, since it is fully based on curvature observables encoded e.g. in infinitesimal holonomies.  Therefore,  even if the eliminativist view is incapable of describing the different,  spatial and curvature-free $A$'s---the different winding numbers,---the integral $\int \ch_A$ could still  have physical significance.
 

A suggestive comparison can be carried out with the observability of the total energy of a  subsystem in classical mechanics. The total energy [$\sim$winding number of a vacuum state] due to one boundary is not a well defined concept, nor a physically meaningful one. 
Nonetheless {\it differences} in energies [$\sim$non-vanishing values of the $\theta_{\text{\tiny YM}}$-term] are meaningful and physically measurable. In classical mechanics there is no absolute concept of energy, but differences in energy are perfectly kosher physical quantities, and one can get by just fine referring solely to such differences. Similarly, one could express all physical quantities solely with the $\theta_{\text{\tiny YM}}$-term without appeal to the individual winding numbers. 
  Indeed, \citet[p. 179]{Healey_book} makes a very similar analogy: 
\begin{quote}
Models related by a ``large'' gauge transformation are characterized by different Chern-Simons numbers, and one might take these to exhibit a difference in the intrinsic properties of the situations they represent. But it is questionalble whether the Chern-Simons number of a gauge-configuration represents an intrinsic property of that configuration, even if a \textit{difference} in Chern-Simons numbers represents an intrinsic \textit{difference} between gauge-configuration. Perhaps Chern-Simons numbers are like velocities in models of special relativity.
\end{quote}


 This observations then underpin the second role of the $ \theta_{\text{\tiny YM}}$-term. That is, gauge theory allows the existence of distinct boundary states (e.g. initial and final states) that are all curvature-free but labelled by different winding numbers. 
These boundary states then represent different choices of initial and final vacua for the theory and the $ \theta_{\text{\tiny YM}}$-term can represent, in a semiclassical (``instanton'') approximation, a transition from one such curvature-free boundary state to a different one \citet{BPS, tHooft_inst}. That is, as we saw, for asymptotically  flat configurations, the Chern number gives a difference between winding numbers, $\int\ch_A=n_+-n_-=:\nu$. If one wants to include configurations with different winding numbers in the path integral, with weight factors $f(\nu)$ for each sector, cluster decomposition of expectation values requires that $f(\nu)=\exp(i\theta \nu)$, where $\theta$ is a free-parameter (cf \citet[p. 456]{WeinbergQFT2}). Thus the inclusion of the  $ \theta_{\text{\tiny YM}}$-term in the Lagrangian corresponds to allowing a superposition of all winding numbers, and the same parameter in the path integral will be included in the superposition of vacuum states.

 Indeed, the impossibility of distinguishing vacuum states with different winding number ($|n\rangle$)  from each other via local observables\footnote{A {\it global} observable that is capable of this is $\CS(A)$.}, jointly with the physical significance of the  \textit{difference} between winding numbers, allows the (formal) introduction of the $\theta$-vacuum state:
\be
|\theta\rangle = \sum_n e^{i \theta n} |n\rangle
\ee
which transforms by a phase under shifts of the winding number.  
Then, each $\theta$-vacuum defines an independent sector of the quantum theory.

 One important point  to observe from this argument, vis \`a vis eliminativism,  is that it is at least a logical possibility to have a representation of $\mathcal{L}_\theta$ in the physics and yet have no way of discerning the  individual winding numbers  entering the $\theta$-vacuum.

  But there are other possibilities. Accounting for certain non-perturbative properties of the quantization of a gauge system \citep[Ch. 3]{Strocchi_SB}, there may be no place  for a physically significant topological $\theta_{\text{\tiny YM}}$-term, and yet   chiral invariance should still be broken without introducing Goldstone bosons. 
Indeed, non-perturbative resolutions of the U$(1)_A$-puzzle  could not resort to the topological properties of the bundle  (since the path integral supposedly has a zero-measure support on continuous fields); rather, they resort to topological properties of the group of local (time-independent) gauge transformations $\mathcal G_3$ that survive the imposition of temporal gauge:

\begin{quote}The  topological invariants [of the group of local gauge transformations $ \mathcal G_3$] defines elements of the center of the local algebra of observables; for Yang-Mills theories such elements [...] are labeled by the winding number  [...] their spectrum labels the factorial representations of the local algebra of observables, the corresponding ground states being the $\theta_{\text{\tiny YM}}$-vacua. They are unstable\footnote{The instability mentioned in this quote is due to the fact that the chiral symmetry acts as what we could call a ``meta-symmetry'' between different $\theta_{\text{\tiny YM}}$-vacua, $\theta \mapsto \theta +\lambda$. Key to the consistency of this formulation is the fact that the limit $\lambda\to 0$ is not properly defined (i.e. the symmetry is not implemented in a weakly continuous way). } under the chiral transformations [..] and therefore chiral transformations are inevitably broken [within each factorial representation (sector) defined by a choice of $\theta_{\text{\tiny YM}}$-vacuum ...] Thus, the topology [of $\G_3$] provides an explanation of chiral symmetry breaking in QCD, without recourse to the instanton semiclassical approximation. \citet[p. 12] {Strocchi_phil}\end{quote}

 In sum, depending on the level of mathematical rigor or the validity of the semiclassical approximation, different accounts of the resolutions of the U$(1)_A$-puzzle can be found in the literature. And even if holonomies are incapable of having a representation of the different connected components of $\G_3$, it could still be true that chiral symmetry is broken without the addition of Goldstone modes, as claimed by \cite{FortGambini}---a claim we will not assess.

Here, we should again emphasize: in this paper, our intent was not to examine the full, non-perturbative quantum picture, nor \citep{FortGambini}'s claims, nor  their relation to \citep{Healey_book}'s, and thus we have refrained from assessing the significance of the $\theta_\text{\tiny YM}$-term in  these respective domains.  Our intent was rather to correct a mistreatment of gauge in the semiclassical picture---irrespective of whether this picture provides a completely satisfactory account of chiral symmetry breaking or not.


\section{Conclusions}\label{sec:conclusions}
\subsection{Summary of our discussion}\label{sec:summary}
On the eliminative view and the gauge-invariant properties of the $\theta_{\text{\tiny YM}}$-term,  \citet[p. 16]{Dougherty_CP} concludes:\begin{quote}
[I] showed that if the eliminative
view were true then the vacuum Yang-Mills $\theta_{\text{\tiny YM}}$-term [\eqref{eq:theta}]
[...]
would lead to inconsistency when integrated over any region [...] By
Stokes' theorem it is a matter of mathematical fact that this integral
coincides with the integral of $\cs_A$.
But this integral varies under large gauge transformations. So if I
were to eliminate gauge from the theory then each configuration
would be assigned contradictory values for the vacuum Yang-Mills
term of the action: one for each class of representative gauge potentials
that differ by a large gauge transformation.
\end{quote}
Our discussion has explained, qualified, and rectified Dougherty's statement.

The $\theta_{\text{\tiny YM}}$-term is manifestly gauge invariant under {\it all} gauge transformations, as shown in section \ref{sec:intro0}. This is just a consequence of the cyclic trace identity and the transformation properties of the curvature---and Stokes' theorem cannot change this  fact. 

Nonetheless, we felt it was important to explain some sources of confusion surrounding  the $\theta_{\text{\tiny YM}}$-term. For instance, it may be expressed as Wess-Zumino integrals on gluing surfaces, and the arguments of these integrals look like gauge transformations; doesn't that indicate their gauge-variance, contrary to the brute fact mentioned above?
 
  This puzzle is straighened once we take into account that the arguments of these integrals  on the gluing surfaces are transition functions, and not gauge transformations, and that in fact, non-trivial transition functions cannot be trivialized by gauge transformations.
 Gauge transformations are smooth, and they are associated to charts of the manifold. These two simple requirements mean gauge transformations cannot affect the value of the integral of $\cs_A$ on the boundary of the manifold, in accord with the invariance of the Chern number.

For configurations that are asymptotically curvature-free, the only way to obtain a non-trivial winding number at the asymptotic boundary requires a non-vanishing curvature for  $A$ in the bulk---$A$ is not a ``pure-gauge'' configuration. That is how the winding number can be represented by the $\theta_{\text{\tiny YM}}$-term---which depends only on the curvature.
In Lorentzian signature (with appropriate boundary conditions at spacelike infinity) this means that transitions over time between winding numbers must be associated with curvature at some point in time.

 Regarding \citet{Dougherty_CP}'s criticism: it invokes a ``size distinction'' by assuming  there is a choice to be made on  whether to accept gauge transformations as acting solely on the boundary of the manifold or not. Moreover, he chains the eliminativist to the more permissive choice, where a restricted gauge transformation, e.g. acting solely at the boundary,  is bona-fide. But no such choice exists: a size-distinction would lead to two different and incompatible notions of gauge.
 A boundary transformation that changes the  (total) winding number  \textit{cannot} be systematically extended to a bulk transformation that sends one solution of the equations of motion to another---as a gauge transformation would---and therefore this transformation cannot be called a symmetry. Indeed, the bulk configuration---including its curvature---has to be changed alongside the change at the boundary necessary for a different winding number.
  
 While it is true that on a manifold with asymptotic boundaries one can nonetheless  use Stokes' theorem to extract  interesting and nontrivial features of the vacuum structure of Yang-Mills theory,  none of these features provide a smoking gun against the eliminative view of gauge, at least in the forms discussed  by \citet[Sec. 6.6]{Healey_book}. 

\subsection{Against eliminativism nonetheless}\label{sec:phil}

Having arrived at the end of this paper, we can  smoke a peace-pipe with Dougherty. As tobacco acceptable to both parties, we notice that the most developed understanding of the solution to the U$(1)_A$-puzzle (i.e. the breaking of chiral symmetry without the introduction of Goldstone bosons), requires the physical significance of structures associated to the existence of the gauge symmetry: be it the role of the fibre bundle topology in the standard semi-classical account, or the role of different connected components of $\mathcal{G}_3$ in the non-perturbative one. In both cases, the arguments bode against any naive implementation of eliminativism.


More broadly, eliminativism of gauge fields is unwarranted for many reasons, some of which we now briefly  summarize. Gauge degrees of freedom simplify mathematical treatments of physical theories by allowing us to write our theories in terms of Lorentz invariant action functionals (and path integrals): there is no available local Hamiltonian or Lagrangian, even in the Abelian case (i.e. electromagnetism) that traffics only in electric and magnetic fields.  Gauge fields are also necessary to maintain certain composition properties: e.g. regionally reduced theories cannot be composed (cf.  \citep{RovelliGauge2013, GomesRiello_new, GomesStudies, Gomes_new, DES_gf}). 
 Moreover, gauge degrees of freedom are introduced to mandate the local Gauss law: action functionals that employ them automatically ensure  both the local Gauss law and  charge conservation. In this sense, gauge degrees of freedom fill an explanatory gap: e.g.  they guarantee  conservation laws along much else. 

Fibre bundles provide a yet deeper explanation of these degrees of freedom through a sort of `internal relationism', in accord with Yang and Mills' original interpretation (cf. Section \ref{sec:intro}). That is, fibre bundles formalize the notion that certain properties that are taken as, in a certain sense,  ``intrinsic'', such as ``being a proton'',\footnote{ Of course this example, which originally motivated Yang and Mills, is meant in the context of the (approximate) isospin symmetry. Otherwise, the electric charge tells protons and neutron apart in an intrinsic manner.}  are in fact relational. Empirical consequences of these relations---Gauss and  conservation laws---follow from this realist-friendly explanation.   
                

\bibliographystyle{apacite} 

\begin{thebibliography}{}

\bibitem [\protect \citeauthoryear {%
A.A.~Belavin%
}{%
A.A.~Belavin%
}{%
{\protect \APACyear {1975}}%
}]{%
BPS}
\APACinsertmetastar {%
BPS}%
\begin{APACrefauthors}%
A.A.~Belavin, A\BPBI S\BPBI Y\BPBI T., A.M.~Polyakov.%
\end{APACrefauthors}%
\unskip\
\newblock
\APACrefYearMonthDay{1975}{}{}.
\newblock
{\BBOQ}\APACrefatitle {Pseudoparticle solutions of the Yang-Mills equations}
  {Pseudoparticle solutions of the yang-mills equations}.{\BBCQ}
\newblock
\APACjournalVolNumPages{Physics Letters B,Volume 59, Issue 1,}{}{}{}.
\PrintBackRefs{\CurrentBib}

\bibitem [\protect \citeauthoryear {%
Bertlmann%
}{%
Bertlmann%
}{%
{\protect \APACyear {1996}}%
}]{%
Bertlmann}
\APACinsertmetastar {%
Bertlmann}%
\begin{APACrefauthors}%
Bertlmann, R.%
\end{APACrefauthors}%
\unskip\
\newblock
\APACrefYear{1996}.
\newblock
\APACrefbtitle {{Anomalies in quantum field theory}} {{Anomalies in quantum
  field theory}}.
\PrintBackRefs{\CurrentBib}

\bibitem [\protect \citeauthoryear {%
DeWitt%
}{%
DeWitt%
}{%
{\protect \APACyear {2003}}%
}]{%
DeWitt_Book2}
\APACinsertmetastar {%
DeWitt_Book2}%
\begin{APACrefauthors}%
DeWitt, B\BPBI S.%
\end{APACrefauthors}%
\unskip\
\newblock
\APACrefYear{2003}.
\newblock
\APACrefbtitle {The Global Approach to Quantum Field Theory, Vol. 2} {The
  global approach to quantum field theory, vol. 2}\ (\BVOL~114).
\newblock
\APACaddressPublisher{}{Clarendon Press, Oxford}.
\PrintBackRefs{\CurrentBib}

\bibitem [\protect \citeauthoryear {%
Dougherty%
}{%
Dougherty%
}{%
{\protect \APACyear {2019}}%
}]{%
Dougherty_CP}
\APACinsertmetastar {%
Dougherty_CP}%
\begin{APACrefauthors}%
Dougherty, J.%
\end{APACrefauthors}%
\unskip\
\newblock
\APACrefYearMonthDay{2019}{}{}.
\newblock
{\BBOQ}\APACrefatitle {Large Gauge Transformations and the Strong CP Problem}
  {Large gauge transformations and the strong cp problem}.{\BBCQ}
\newblock
\APACjournalVolNumPages{Studies in the History and Philosophy of Modern
  Physics, vol 37, 2020}{}{}{}.
\PrintBackRefs{\CurrentBib}

\bibitem [\protect \citeauthoryear {%
Earman%
}{%
Earman%
}{%
{\protect \APACyear {2004}}%
}]{%
Earman_sym}
\APACinsertmetastar {%
Earman_sym}%
\begin{APACrefauthors}%
Earman, J.%
\end{APACrefauthors}%
\unskip\
\newblock
\APACrefYearMonthDay{2004}{}{}.
\newblock
{\BBOQ}\APACrefatitle {Laws, Symmetry, and Symmetry Breaking: Invariance,
  Conservation Principles, and Objectivity} {Laws, symmetry, and symmetry
  breaking: Invariance, conservation principles, and objectivity}.{\BBCQ}
\newblock
\APACjournalVolNumPages{Philosophy of Science}{71}{5}{1227--1241}.
\newblock
\begin{APACrefURL} \url{https://www.jstor.org/stable/10.1086/428016}
  \end{APACrefURL}
\PrintBackRefs{\CurrentBib}

\bibitem [\protect \citeauthoryear {%
Fort%
\ \BBA {} Gambini%
}{%
Fort%
\ \BBA {} Gambini%
}{%
{\protect \APACyear {2000}}%
}]{%
FortGambini}
\APACinsertmetastar {%
FortGambini}%
\begin{APACrefauthors}%
Fort, H.%
\BCBT {}\ \BBA {} Gambini, R.%
\end{APACrefauthors}%
\unskip\
\newblock
\APACrefYearMonthDay{2000}{}{}.
\newblock
{\BBOQ}\APACrefatitle {U(1) Puzzle and the Strong CP Problem from a Holonomy
  Formulation Perspective} {U(1) puzzle and the strong cp problem from a
  holonomy formulation perspective}.{\BBCQ}
\newblock
\APACjournalVolNumPages{International Journal of Theoretical Physics, Vol. 39,
  No. 2}{}{}{}.
\PrintBackRefs{\CurrentBib}

\bibitem [\protect \citeauthoryear {%
Gomes%
}{%
Gomes%
}{%
{\protect \APACyear {2019}}%
{\protect \APACexlab {{\protect \BCnt {1}}}}}]{%
GomesStudies}
\APACinsertmetastar {%
GomesStudies}%
\begin{APACrefauthors}%
Gomes, H.%
\end{APACrefauthors}%
\unskip\
\newblock
\APACrefYearMonthDay{2019{\protect \BCnt {1}}}{}{}.
\newblock
{\BBOQ}\APACrefatitle {Gauging the boundary in field-space} {Gauging the
  boundary in field-space}.{\BBCQ}
\newblock
\APACjournalVolNumPages{Studies in History and Philosophy of Science Part B:
  Studies in History and Philosophy of Modern Physics}{}{}{}.
\newblock
\begin{APACrefURL}
  \url{http://www.sciencedirect.com/science/article/pii/S1355219818302144}
  \end{APACrefURL}
\newblock
\begin{APACrefDOI} \doi{https://doi.org/10.1016/j.shpsb.2019.04.002}
  \end{APACrefDOI}
\PrintBackRefs{\CurrentBib}

\bibitem [\protect \citeauthoryear {%
Gomes%
}{%
Gomes%
}{%
{\protect \APACyear {2019}}%
{\protect \APACexlab {{\protect \BCnt {2}}}}}]{%
Gomes_new}
\APACinsertmetastar {%
Gomes_new}%
\begin{APACrefauthors}%
Gomes, H.%
\end{APACrefauthors}%
\unskip\
\newblock
\APACrefYearMonthDay{2019{\protect \BCnt {2}}}{}{}.
\newblock
{\BBOQ}\APACrefatitle {Holism as the significance of gauge symmetries} {Holism
  as the significance of gauge symmetries}.{\BBCQ}
\newblock

\PrintBackRefs{\CurrentBib}

\bibitem [\protect \citeauthoryear {%
Gomes%
}{%
Gomes%
}{%
{\protect \APACyear {2020}}%
}]{%
DES_gf}
\APACinsertmetastar {%
DES_gf}%
\begin{APACrefauthors}%
Gomes, H.%
\end{APACrefauthors}%
\unskip\
\newblock
\APACrefYearMonthDay{2020}{}{}.
\newblock
{\BBOQ}\APACrefatitle {Gauge-invariance and the direct empirical significance
  of symmetries} {Gauge-invariance and the direct empirical significance of
  symmetries}.{\BBCQ}
\newblock
\APACjournalVolNumPages{(to appear)}{}{}{}.
\PrintBackRefs{\CurrentBib}

\bibitem [\protect \citeauthoryear {%
Gomes%
\ \BBA {} Riello%
}{%
Gomes%
\ \BBA {} Riello%
}{%
{\protect \APACyear {2019}}%
}]{%
GomesRiello_new}
\APACinsertmetastar {%
GomesRiello_new}%
\begin{APACrefauthors}%
Gomes, H.%
\BCBT {}\ \BBA {} Riello, A.%
\end{APACrefauthors}%
\unskip\
\newblock
\APACrefYearMonthDay{2019}{}{}.
\newblock
{\BBOQ}\APACrefatitle {{Quasilocal degrees of freedom in Yang-Mills theory}}
  {{Quasilocal degrees of freedom in Yang-Mills theory}}.{\BBCQ}
\newblock
\APACjournalVolNumPages{Forthcoming in SciPost}{}{}{}.
\PrintBackRefs{\CurrentBib}

\bibitem [\protect \citeauthoryear {%
Hawking%
\ \BBA {} Ellis%
}{%
Hawking%
\ \BBA {} Ellis%
}{%
{\protect \APACyear {1975}}%
}]{%
hawking1975large}
\APACinsertmetastar {%
hawking1975large}%
\begin{APACrefauthors}%
Hawking, S\BPBI W.%
\BCBT {}\ \BBA {} Ellis, G\BPBI F\BPBI R.%
\end{APACrefauthors}%
\unskip\
\newblock
\APACrefYear{1975}.
\newblock
\APACrefbtitle {The Large Scale Structure of Space-Time (Cambridge Monographs
  on Mathematical Physics)} {The large scale structure of space-time (cambridge
  monographs on mathematical physics)}.
\newblock
\APACaddressPublisher{}{Cambridge University Press}.
\newblock
\begin{APACrefURL}
  \url{http://www.amazon.com/Structure-Space-Time-Cambridge-Monographs-Mathematical/dp/0521099064}
  \end{APACrefURL}
\PrintBackRefs{\CurrentBib}

\bibitem [\protect \citeauthoryear {%
Healey%
}{%
Healey%
}{%
{\protect \APACyear {2007}}%
}]{%
Healey_book}
\APACinsertmetastar {%
Healey_book}%
\begin{APACrefauthors}%
Healey, R.%
\end{APACrefauthors}%
\unskip\
\newblock
\APACrefYear{2007}.
\newblock
\APACrefbtitle {Gauging What's Real: The Conceptual Foundations of Gauge
  Theories} {Gauging what's real: The conceptual foundations of gauge
  theories}.
\newblock
\APACaddressPublisher{}{Oxford University Press}.
\PrintBackRefs{\CurrentBib}

\bibitem [\protect \citeauthoryear {%
Kobayashi%
\ \BBA {} Nomizu%
}{%
Kobayashi%
\ \BBA {} Nomizu%
}{%
{\protect \APACyear {1963}}%
}]{%
kobayashivol1}
\APACinsertmetastar {%
kobayashivol1}%
\begin{APACrefauthors}%
Kobayashi, S.%
\BCBT {}\ \BBA {} Nomizu, K.%
\end{APACrefauthors}%
\unskip\
\newblock
\APACrefYear{1963}.
\newblock
\APACrefbtitle {Foundations of differential geometry. {V}ol {I}} {Foundations
  of differential geometry. {V}ol {I}}.
\newblock
\APACaddressPublisher{}{Interscience Publishers, a division of John Wiley \&
  Sons, New York-Lond on}.
\PrintBackRefs{\CurrentBib}

\bibitem [\protect \citeauthoryear {%
Nakahara%
}{%
Nakahara%
}{%
{\protect \APACyear {2003}}%
}]{%
Nakahara}
\APACinsertmetastar {%
Nakahara}%
\begin{APACrefauthors}%
Nakahara, M.%
\end{APACrefauthors}%
\unskip\
\newblock
\APACrefYear{2003}.
\newblock
\APACrefbtitle {Geometry, Topology and Physics} {Geometry, topology and
  physics}.
\newblock
\APACaddressPublisher{}{Institute of Physics}.
\PrintBackRefs{\CurrentBib}

\bibitem [\protect \citeauthoryear {%
Rovelli%
}{%
Rovelli%
}{%
{\protect \APACyear {2014}}%
}]{%
RovelliGauge2013}
\APACinsertmetastar {%
RovelliGauge2013}%
\begin{APACrefauthors}%
Rovelli, C.%
\end{APACrefauthors}%
\unskip\
\newblock
\APACrefYearMonthDay{2014}{}{}.
\newblock
{\BBOQ}\APACrefatitle {{Why Gauge?}} {{Why Gauge?}}{\BBCQ}
\newblock
\APACjournalVolNumPages{Found. Phys.}{44}{1}{91-104}.
\newblock
\begin{APACrefDOI} \doi{10.1007/s10701-013-9768-7} \end{APACrefDOI}
\PrintBackRefs{\CurrentBib}

\bibitem [\protect \citeauthoryear {%
Schäfer%
\ \BBA {} Shuryak%
}{%
Schäfer%
\ \BBA {} Shuryak%
}{%
{\protect \APACyear {1998}}%
}]{%
CP_Schafer}
\APACinsertmetastar {%
CP_Schafer}%
\begin{APACrefauthors}%
Schäfer, T.%
\BCBT {}\ \BBA {} Shuryak, E\BPBI V.%
\end{APACrefauthors}%
\unskip\
\newblock
\APACrefYearMonthDay{1998}{}{}.
\newblock
{\BBOQ}\APACrefatitle {{Instantons in QCD}} {{Instantons in QCD}}.{\BBCQ}
\newblock
\APACjournalVolNumPages{Rev. Mod. Phys.}{70}{}{323--426}.
\newblock
\begin{APACrefDOI} \doi{10.1103/RevModPhys.70.323} \end{APACrefDOI}
\PrintBackRefs{\CurrentBib}

\bibitem [\protect \citeauthoryear {%
Strocchi%
}{%
Strocchi%
}{%
{\protect \APACyear {2015}}%
}]{%
Strocchi_phil}
\APACinsertmetastar {%
Strocchi_phil}%
\begin{APACrefauthors}%
Strocchi, F.%
\end{APACrefauthors}%
\unskip\
\newblock
\APACrefYearMonthDay{2015}{}{}.
\newblock
{\BBOQ}\APACrefatitle {{Symmetries, Symmetry Breaking, Gauge Symmetries}}
  {{Symmetries, Symmetry Breaking, Gauge Symmetries}}.{\BBCQ}
\newblock

\PrintBackRefs{\CurrentBib}

\bibitem [\protect \citeauthoryear {%
Strocchi%
}{%
Strocchi%
}{%
{\protect \APACyear {2019}}%
}]{%
Strocchi_SB}
\APACinsertmetastar {%
Strocchi_SB}%
\begin{APACrefauthors}%
Strocchi, F.%
\end{APACrefauthors}%
\unskip\
\newblock
\APACrefYear{2019}.
\newblock
\APACrefbtitle {Symmetry breaking in the standard model: a non-perturbative
  outlook} {Symmetry breaking in the standard model: a non-perturbative
  outlook}.
\newblock
\APACaddressPublisher{}{Springer lecture notes}.
\PrintBackRefs{\CurrentBib}

\bibitem [\protect \citeauthoryear {%
't Hooft%
}{%
't Hooft%
}{%
{\protect \APACyear {1976}}%
}]{%
tHooft_inst}
\APACinsertmetastar {%
tHooft_inst}%
\begin{APACrefauthors}%
't Hooft, G.%
\end{APACrefauthors}%
\unskip\
\newblock
\APACrefYearMonthDay{1976}{Dec}{}.
\newblock
{\BBOQ}\APACrefatitle {Computation of the quantum effects due to a
  four-dimensional pseudoparticle} {Computation of the quantum effects due to a
  four-dimensional pseudoparticle}.{\BBCQ}
\newblock
\APACjournalVolNumPages{Phys. Rev. D}{14}{}{3432--3450}.
\newblock
\begin{APACrefURL} \url{https://link.aps.org/doi/10.1103/PhysRevD.14.3432}
  \end{APACrefURL}
\newblock
\begin{APACrefDOI} \doi{10.1103/PhysRevD.14.3432} \end{APACrefDOI}
\PrintBackRefs{\CurrentBib}

\bibitem [\protect \citeauthoryear {%
Weinberg%
}{%
Weinberg%
}{%
{\protect \APACyear {2005}}%
}]{%
WeinbergQFT2}
\APACinsertmetastar {%
WeinbergQFT2}%
\begin{APACrefauthors}%
Weinberg, S.%
\end{APACrefauthors}%
\unskip\
\newblock
\APACrefYear{2005}.
\newblock
\APACrefbtitle {The Quantum Theory of Fields. Volume 2. Modern Applications}
  {The quantum theory of fields. volume 2. modern applications}.
\newblock
\APACaddressPublisher{}{Cambridge Univ. Press}.
\PrintBackRefs{\CurrentBib}

\bibitem [\protect \citeauthoryear {%
Yang%
\ \BBA {} Mills%
}{%
Yang%
\ \BBA {} Mills%
}{%
{\protect \APACyear {1954}}%
}]{%
YangMills}
\APACinsertmetastar {%
YangMills}%
\begin{APACrefauthors}%
Yang, C\BPBI N.%
\BCBT {}\ \BBA {} Mills, R\BPBI L.%
\end{APACrefauthors}%
\unskip\
\newblock
\APACrefYearMonthDay{1954}{Oct}{}.
\newblock
{\BBOQ}\APACrefatitle {Conservation of Isotopic Spin and Isotopic Gauge
  Invariance} {Conservation of isotopic spin and isotopic gauge
  invariance}.{\BBCQ}
\newblock
\APACjournalVolNumPages{Phys. Rev.}{96}{}{191--195}.
\newblock
\begin{APACrefURL} \url{https://link.aps.org/doi/10.1103/PhysRev.96.191}
  \end{APACrefURL}
\newblock
\begin{APACrefDOI} \doi{10.1103/PhysRev.96.191} \end{APACrefDOI}
\PrintBackRefs{\CurrentBib}

\end{thebibliography}

\end{document}